\documentclass{appolb}
\usepackage{epsfig}

\newcommand{\sqrtsNN}{\mbox{$\sqrt{\mathrm{s}_{_{NN}}}$} }
\newcommand{\vtwo}{$v_{2}$ }

\newcommand{\ks}{${K}^{0}_{S}$ }
\newcommand{\lam}{$\Lambda$ }

\def \GeVc {\mbox{$\mathrm{GeV}/c$}}

\def \auau  {$\mathrm{Au + Au}$ }

\begin{document}
\title{The elliptic flow in Au+Au collisions at \sqrtsNN = 7.7, 11.5 and 39 GeV at STAR%
\thanks{Presented at the conference `Strangeness in Quark Matter 2011', Cracow, Poland, September 18-24, 2011}%
}
\author{Shusu Shi (for the STAR collaboration)
\address[a]{Institute of Particle Physics, Central China
Normal University, Wuhan, Hubei, 430079, China}
\address[b]{The Key Laboratory of Quark and Lepton Physics (Central China Normal
University), Ministry of Education, Wuhan, Hubei, 430079, China}
}
\maketitle
\begin{abstract}
We present elliptic flow, $v_2$, measurements for
charged and identified particles at midrapidity in Au+Au collisions at \sqrtsNN = 7.7, 11.5 and 39 GeV at STAR.
We compare the inclusive charged hadron $v_2$ to those
from high energies at RHIC (\sqrtsNN = 62.4 and 200 GeV), at LHC
(\sqrtsNN = 2.76 TeV). A significant difference in $v_2$ between baryons and
anti-baryons is observed and the difference increases with decreasing beam energy.
We observed the $v_2$ of $\phi$ meson is systematically lower than other particles in
Au+Au collisions at \sqrtsNN = 11.5 GeV.
\end{abstract}

\section{Introduction}
The main physics motivation for the Beam Energy Scan at RHIC-STAR experiment is searching for the phase boundary and
critical point predicted by QCD theory. The elliptic flow, $v_2$, which is generated by the initial anisotropy in the
coordinate space, is defined by
\begin{equation} v_{2}=\langle\cos2(\phi-\Psi_{R})\rangle
\end{equation}, where $\phi$ is azimuthal angle of an outgoing particle,
$\Psi_{R}$ is the azimuthal angle of the impact parameter, and
angular brackets denote an average over many particles and events.
Due to the self-quenching effect, it is sensitive to the early stage of the heavy ion collisions~\cite{review}.
The number of constituent quark (NCQ) scaling observed in the top energy of RHIC Au+Au and Cu+Cu collisions~\cite{starklv2, XiOmega, flowcucu} reflects that
the collectivity has been built up at the partonic stage. Especially, the NCQ scaling for multi-strange hadrons, $\phi$ ($s\overline{s}$) and $\Omega$ ($sss$)
supports the deconfinement and partonic collectivity picture~\cite{statphi, phiomega}.
A study based on a multi-phase transport model (AMPT) indicates the NCQ scaling is related to the degrees of freedom in the system~\cite{AMPTNCQ}.
The scaling and no scaling in $v_2$ reflects the partonic and hadronic degrees of freedom respectively.
The importance of $\phi$ meson has been
emphasized, where the $\phi$ meson $v_2$ could be small or zero without
partonic phase~\cite{phiBES}.
Thus, the measurements of elliptic flow with the Beam Energy Scan data offer us the opportunity to investigate the QCD phase boundary.

\begin{figure*}[ht]
\vskip 0cm
\begin{center} \includegraphics[width=0.7\textwidth]{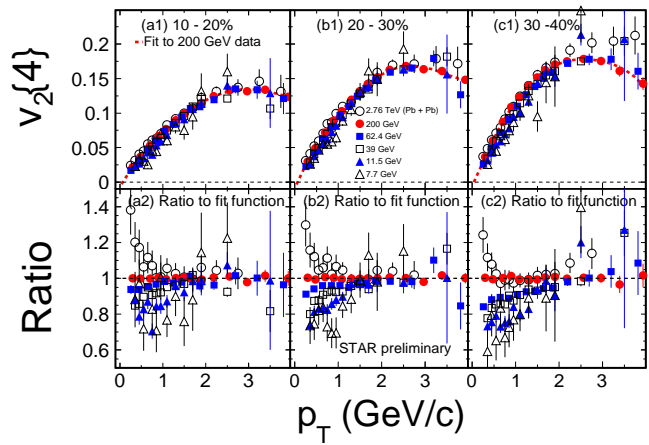}\end{center}
\caption{(Color online) The top panels show the $v_2\{4\}$ vs. $p_T$ at midrapidity for various beam energies (\sqrtsNN = 7.7 GeV to 2.76 TeV).
The results for \sqrtsNN = 7.7 to 200 GeV are for \auau collisions and those for 2.76 TeV are for Pb + Pb collisions.
The red line shows the fit to the results from \auau collisions at \sqrtsNN = 200 GeV. The bottom panels show
the ratio of $v_2\{4\}$ vs. $p_T$ for all \sqrtsNN with respect to this fitted line. The results are shown for three collision centrality classes:
$10 - 20\%$ (a1), $20 - 30\%$ (b1) and $30 - 40\%$ (c1).}
\label{v2_4_pt_beam_energy}
\end{figure*}

In this proceedings, we present the \vtwo results of charged and identified hadrons
by the STAR experiment from Au+Au collisions in \sqrtsNN = 7.7, 11.5 and 39 GeV.
STAR's Time Projection Chamber (TPC)~\cite{STARtpc} is used as the
main detector for event plane determination. The
centrality was determined by the number of tracks from the pseudorapidity region
$|\eta|\le 0.5$. The particle identification for $\pi^{\pm}$, $K^{\pm}$ and $p~(\overline{p})$ is
achieved via the energy loss in the TPC and the time of flight information from the multi-gap resistive plate chamber detector~\cite{STARtof}.
Strange hadrons are reconstructed with the decay channels:
\ks $\rightarrow \pi^{+} + \pi^{-}$, $\phi \rightarrow K^{+} +
K^{-}$, \lam $\rightarrow p + \pi^{-}$ ($\overline{\Lambda}
\rightarrow \overline{p} + \pi^{+}$),
and $\Xi^{-} \rightarrow$ \lam $+\ \pi^{-}$ ($\overline{\Xi}^{+}
\rightarrow$ $\overline{\Lambda}$+\ $\pi^{+}$)). The detailed description of the
procedure can be found in Refs.~\cite{starklv2, XiOmega,klv2_130GeV}.
The event plane method~\cite{v2Methods1} and cumulant method~\cite{cumulant1, cumulant2} are used for the $v_{2}$ measurement.

\section{Results and Discussions}

\begin{figure*}[ht]
\vskip 0cm
\begin{center}\includegraphics[width=0.6\textwidth]{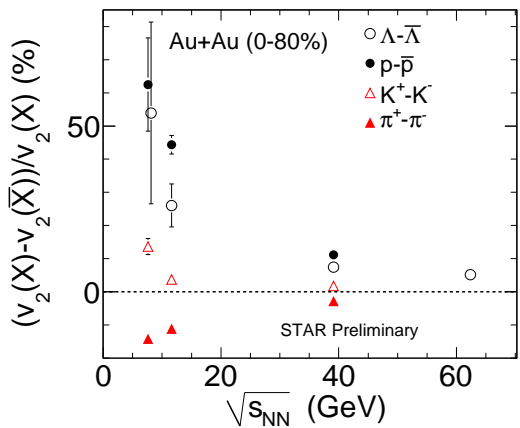}\end{center}
\caption{(Color online) The difference of $v_2$ for particles and anti-particles ($v_{2}(X)-v_{2}(\overline{X})$)
divided by particle $v_2$ ($v_{2}(X)$) as a function of beam energy in Au+Au collisions (0-80\%).}
\label{v2_diff}
\end{figure*}

\begin{figure*}[ht]
\vskip 0cm
\includegraphics[width=1.\textwidth]{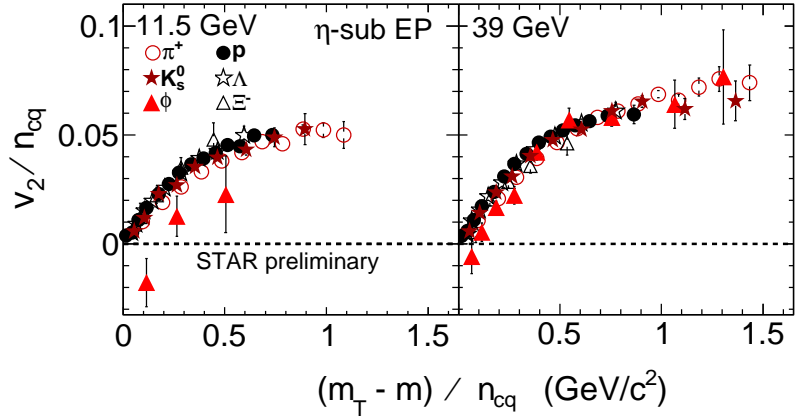}
\caption{(Color online) The number of constituent quark ($n_{\rm cq}$) scaled $v_2$ as a function of transverse kinetic energy over $n_{\rm cq}$
($(m_T - m_0)/n_{\rm cq}$) for various identified particles in Au+Au (0-80\%) collisions at \sqrtsNN = 11.5 and 39 GeV.}
\label{ncq}
\end{figure*}

The Beam Energy Scan data from RHIC-STAR experiment, offer a opportunity to study the beam energy dependence of $v_2$
in a wide range of beam energy.
Figure~\ref{v2_4_pt_beam_energy} shows the results of transverse momentum ($p_T$) dependence of
$v_2\{4\}$ for charged hadrons from  \sqrtsNN = 7.7 GeV to 2.76 TeV
in $10 - 20\%$ (a1), $20 - 30\%$ (b1) and $30 - 40\%$ (c1) centrality bins, where the ALICE results in Pb + Pb collisions at $\sqrt{s_{NN}}$ = 2.76 TeV are taken from~\cite{alicev2}.
At low $p_T$ ($p_T < 2~\GeVc$), the $v_2$ values increases with increase in beam energy. Beyond
$p_T = 2~\GeVc$ the $v_2$ results show comparable values with in statistic errors.
There is no saturation signal of $v_2$ up to collisions at \sqrtsNN = 2.76 TeV.
Figure~\ref{v2_diff} shows excitation function for the relative difference of $v_2$ between particles and
 anti-particles.
Here, to reduce the non-flow effect, the $\eta$-sub event plane method is used to calculate $v_2$.
The $\eta$-sub event plane method is similar to the event plane method, except one
defines the flow vector for each particle based on particles
measured in the opposite hemisphere in pseudorapidity.
An $\eta$ gap of $|\eta| < 0.05$ is used between negative/positive $\eta$ sub-event
to guarantee that non-flow effects are reduced by enlarging the separation
between the correlated particles.
The difference for baryon and anti-baryon (protons and $\Lambda$s) could be observed
from 7.7 to 62.4 GeV. The difference of $v_2$ for baryons is
within 10\% at \sqrtsNN = 39 and 62.4 GeV, while a significant
difference is observed below \sqrtsNN = 39 GeV. For example,
the difference of protons versus anti-protons is around 60\%. There is no obvious difference for
$\pi^{+}$ versus $\pi^{-}$ (within 3\%) and $K^{+}$ versus $K^{-}$ (within 2\%)
when \sqrtsNN = 39 GeV.
As the decrease of beam energy,
$\pi^{+}$ versus $\pi^{-}$ and $K^{+}$ versus $K^{-}$ start to show the difference.
The $v_2$ of $\pi^{-}$ is larger than that of
$\pi^{+}$ and the $v_2$ of $K^{+}$ is larger than that of $K^{-}$. This difference between particles and
anti-particles might be due to the baryon transport effect to midrapidity~\cite{transporteffect} or absorption effect in the hadronic stage.
The results could indicate the hadronic interaction become more dominant in lower beam energy.
The immediate
consequence of the significant difference between baryon and
anti-baryon $v_2$ is that the NCQ scaling is broken between particles and
anti-particles at \sqrtsNN = 7.7 and 11.5 GeV.
The transverse momentum dependence of $v_2$ for the selected identified particles is shown in Fig.~\ref{ncq}.
The $v_2$ and $m_{T}-m_{0}$ has been divided by number of constituent quark in each hadron. In Au+Au collisions at \sqrtsNN = 39 GeV,
the similar scaling behavior at \sqrtsNN =200 GeV is observed. Especially,
the $\phi$ mesons which is insensitive to the later hadronic rescatterings follows the same trend of other particles. It suggests that the
partonic collectivity has been built up in collisions at \sqrtsNN = 39 GeV.
However, the $v_2$ for $\phi$ mesons falls off from other
particles at \sqrtsNN = 11.5 GeV. The mean deviation to the $v_2$ pions is 2.6 $\sigma$.
It indicates that the hadronic interaction
are dominant in collisions at \sqrtsNN = 11.5 GeV.

\section{Summary}
In summary, we present the $v_2$ measurement for charged hadrons and identified hadrons in Au+Au collisions at
 \sqrtsNN= 7.7, 11.5 and 39 GeV. The magitude of $v_2$ increases as increasing of the beam energy
from 7.7 GeV to 2.76 TeV. The difference between the $v_2$ of particles and anti-particles is observed.
The baryon and anti-baryon show significant difference at \sqrtsNN = 7.7 and 11.5 GeV.  
The ongoing analysis with 19.6 and 27 GeV data collected in 2011 will fill the gap between 11.5 and 39 GeV.
The pions and kaons are almost consistent at \sqrtsNN = 39 GeV. This difference
increase with decreasing of the beam energy. The $v_2$ of $\phi$ meson falls off from other particles in
collisions at \sqrtsNN = 11.5 GeV. Experimental data suggests the hadronic interactions are dominant when
\sqrtsNN $\leq$ 11.5 GeV.

\section{Acknowledgments}
This work was supported in part by the National Natural Science Foundation of China under grant No. 11105060, 10775060 and 11135011.

\end{document}